\documentclass{elsart}
\usepackage{amssymb}
\usepackage{amsmath}
 \usepackage[dvips]{graphicx}
\journal{   ArXiv \& Hal\qquad\qquad\qquad\qquad\qquad\qquad\qquad\qquad\qquad\qquad\qquad\qquad\qquad\qquad\qquad\qquad}
\input{tcilatex}
\begin{document}

\begin{frontmatter}

\title{Utilization of the second gradient theory in continuum mechanics to study
 motions and thermodynamics of liquid-vapor interfaces}

\author{Henri Gouin},

\ead{henri.gouin@univ-cezanne.fr}

\address{   University of Aix-Marseille \& U.M.R. C.N.R.S.
6181\\
 Av. Escadrille Normandie-Niemen, Box 322, 13397 Marseille Cedex 20, France.  }
\medskip
\emph{ \textbf{\footnotesize Revisited and completed version of Physicochemical Hydrodynamics,\\
 Series B, Physics,  vol. 174, 1986, p.p. 667-682}}

\address{}

\begin{abstract}
A thermomechanical model of continuous fluid media based on second gradient theory
is used to study motions in liquid-vapor interfaces.  At equilibrium,
the model is shown to be  equivalent to mean-field molecular theories of capillarity.
In such fluids, conservative motions  verify  first integrals that lead  to
Kelvin  circulation theorems and potential equations. The dynamical surface
tension of liquid-vapor interfaces is deduced from viscous fluid equations.  The result provides and explains the Marangoni effect.

\end{abstract}


\end{frontmatter}

\section{Introduction}

Usually, liquid-vapor interfaces are represented by   material
surfaces endowed with an energy related to the Laplace surface tension. Several
studies conducted in the fields of fluid mechanics and thermodynamics
represent the interface as a surface separating two media and having its own energy density and
characteristic behaviors \cite%
{Defay,Sanfeld,Emschwiller,Davis,Landau,Sen,Ono,Levitch}.  When working far from the
critical temperature, detailed
theoretical or experimental studies show that the capillary layer thickness ranges over a few
molecular beams \cite{Emschwiller,Levitch,Rocard}. For   fluid conditions defined by
 temperature and density, molecular models such as
those used in gas kinetic theory lead to state laws associated with
non-convex internal energies as the one associated with the van der Waals model for the pressure \cite%
{Rocard,Cahn,Dunn,Brin,Rowlinson,Hohenberg}. Such models provide a direct verification of the Maxwell rule applied to isothermal
liquid-vapor transition \cite{Aifantis}; nonetheless, they present two disadvantages:
\newline
First, for density values ranging between the vapor and liquid bulk densities,
the pressure may become negative; however, traction created by simple
physical experiments can lead to such pressure values \cite{Rocard,Brin,Bruhat}.\newline
Second, in the field between vapor and liquid, the internal energy distribution cannot be
represented by a convex surface associated with the  density and
entropy and seems to be inconsistent with the existence of
equilibrium state.\newline
To overcome these disadvantages, the thermodynamical investigation uses a
plane domain in place of the portion corresponding to the  non-convex internal
energy. The fluid can no longer be considered as a continuous medium. The interface is represented by a material surface with a null
thickness. In such a case, the only possible representation of the dynamic
behavior of the interface is  a discontinuous surface and its
essential structure remains unknown.  At  equilibrium, the
disadvantages can be eliminated by appropriately modifying the stress tensor
of the capillary layer  which  is then expressed in an anisotropic form.
As a consequence, the energy of the continuous medium must be modified \cite{Cahn,Casal 1,Bongiorno}.

If the fluid interface must be represented as a continuous medium, how can
the stress tensor of the capillary layer be written in a dynamic expression \cite{Davis}?
In previous articles, we proposed a dynamic theory named   \emph{internal capillarity} based on the second
gradient of deformation of the medium \cite{Casal 2,Casal 3,Gouin 1}. Elaborated in the sixties for the static
case, the theory points out the advantage of using a three-dimensional
approach of the capillarity in a continuous mechanical model \cite{Casal 1,Casal 4}. The
deduced equations of equilibrium provide a satisfying representation of
isothermal liquid-vapor equilibrium states. The approach is not new, and  dates back to van der Waals and Korteweg \cite{van der Waals,Korteweg}. Cahn and Hilliard again used
such an approach in 1959 in reference to the free energy \cite{Cahn}, and went back to a
Landau-Ginsburg model \cite{Hohenberg}. The representation proposed in the present study is
based on the notion of internal energy which is more useful when the
temperature is not uniform. The apparent contradiction between the Korteweg
classical stress theory and the Clausius-Duhem inequality makes the dynamic study of
phase transformation  more difficult \cite{Gurtin}, but the use
of the capillary fluid model rectifies this apparent anomaly \cite{Dunn,Casal 2}. In the more general
case of internal capillarity, by representing energy in terms of second
gradient of deformation  \cite{Germain 1} and by applying a simple algebraic identity, it is possible to draw
a relationship between energy equation, motion equation, mass conservation
equation and entropy \cite{Casal 2}. For a conservative fluid, an additional term with a
heat flux dimension must be introduced into the energy equation. In the case of
viscous fluid, previous results provide a set of equations which do not
modify the Clausius-Duhem inequality and make them compatible with
the second law of thermodynamics. In the non-viscous case, classical fluid
motions and motions of fluids endowed with internal capillarity reveal a
common structure that induces a  thermodynamic form of equation of motion and potential
equations \cite{Casal 1,Casal 5,Seliger}. This leads to the same classification of motions,
generalization of Kelvin theorems,  Crocco-Vaszonyi equation and
first integrals   conserved across the interfaces \cite{Gouin 2,Gouin 3}. By
representing the internal energy as a function of entropy, density, density
gradient and by using a single constant $\lambda$, the
resulting equations are thereby identical to those obtained with molecular
models of mean field theory in the isothermal case \cite{Rocard}.

For a surface area relatively large with respect to   the
capillary layer thickness, the surface tension is calculated by  integration
throughout the interface. Lengthwise,
 the surface tension is not constant and depends on the
dynamical distribution of density and temperature. This dynamical
distribution  based on equations of motion, calls upon a Navier-Stokes
viscosity. For interfaces in   isothermal equilibrium, the results are
classical \cite{Rocard,Cahn,Brin,Rowlinson,Bongiorno}. The study of motion in the interface without mass transfer needs
the surface tension gradient and the velocity gradients associated with the
dynamic viscosity. The ensuing Marangoni effect has been interpreted by
using a limit analysis wherein the approximate quantities correspond to the
physical dimensions of the interface. When the  mass flow across the
interface is not null, a general dynamic form of the Laplace equation is given
  \cite{Gouin 1}. The method herein presented  is completely different from classical
calculations based on the balance equations established for both sides of a
discontinuous surface, which takes account of density variations only as
a difference across the interface \cite{Ishi}. In the particular case of isothermal
liquid-vapor equilibria, an invariant integral of motions compatible with
the interface coincides with a generalization of the rule advanced by
Maxwell, associated with phase transitions \cite{Aifantis}.

The model of a viscous fluid endowed with internal capillarity is therefore
substantiated by the consequences verified in both equilibrium states and
non-isothermal motions. The model provides a better understanding of the
behavior of liquid-vapor interfaces in motion. It at least gives  a partial
answer to the question: \emph{Is the fluid at the interface rigid or moving ?} \cite{Birkhoff}
and proposes a theory which takes   account of the stress tensor and
dynamics in the structure of the interface \cite{Davis}. The resulting
behaviors are not the  laws governing classical fluids since
they include an anisotropic stress tensor in the momentum equation and an
additional heat flux term in the energy equation.

 \newpage

\section{Equations of motion for a fluid endowed with internal capillarity}

\subsection{Case of conservative fluids}

The second gradient theory \cite{Germain 1}, conceptually more straightforward than the Laplace
theory, can be used to construct a theory of capillarity. In the present
text the only addition with respect to compressible fluids is an internal mass energy  which is a
function of density $\rho$, entropy $s$ and also $\func{grad}\rho $.
The specific internal   energy $\varepsilon $ characterizes both the compressibility and
capillarity properties of the fluid \cite{Pratz},
\begin{equation*}
\varepsilon =f(\rho ,s,\beta ) ,  \label{1}
\end{equation*}%
where
\begin{equation*}
\beta =(\func{grad}\rho )^{2}.
\end{equation*}%
The equation of motion for a conservative fluid writes
\begin{equation}
\rho\, \mathbf{a} =\func{div}\mathbf{\mathbf{\sigma}} -\rho \, \func{grad}%
\Omega , \label{2}
\end{equation}%
where $\mathbf{a} $ denotes the acceleration vector, $\Omega $ the
extraneous force potential and $\mathbf{\mathbf{\sigma}} $ the general
stress tensor
\begin{equation}
\mathbf{\mathbf{\sigma}}=-p\,\mathbf{I}-\lambda\,(\func{grad}\rho )(\func{grad}\rho
)^{T}  \label{3a}
\end{equation}%
 or
\begin{equation*}
\begin{array}{ccc}
{\sigma} _{ij}=-p\,\delta _{ij}-\lambda\,\rho _{,i}\rho _{,j}\,, \quad i,j\in
\{1,2,3\} &  &
\end{array}%
\end{equation*}%
where%
\begin{equation*}
\lambda =2\,\rho\, \varepsilon '_{\beta } \label{4}
\end{equation*}%
and
\begin{equation}
p=\rho ^{2}\varepsilon '_{\rho } -\func{div}(\lambda\,\func{grad}\rho ) \label{sigma}
\end{equation}%
It should be noted that:
\begin{equation}
\theta =\varepsilon _{s}^{\prime } \label{temperature}
\end{equation}
is the Kelvin temperature expressed as a function of $\rho$, $s$ and $\beta$.

\emph{Proof}:
The equation of motion is obtained  by using the
virtual work principle. The virtual displacement, denoted by $\delta \textbf{x}$, has been
defined by J. Serrin   (\cite{Serrin}, p. 145). For a fluid endowed with internal
capillarity, the virtual work principle (\cite{Serrin}, iv, section 14) is stated as
follows:
the motion of a fluid is such that:
\begin{equation*}
\delta \left(\int_{V}\rho\, (\varepsilon +\Omega )\, dv\right)-\int_{ V}\rho\, \mathbf{%
a}\, \delta \textbf{x}\, dv =0,
\end{equation*}%
where $V$ is an arbitrary material volume. The variation of entropy (\cite{Serrin}, p.
148) must satisfied the condition $\delta s=0$; while taking  account
of
\begin{equation*}
\delta \left(\frac{\partial \rho }{\partial \textbf{x}}\right)=\frac{\partial \delta \rho }{%
\partial \textbf{x}}-\frac{\partial \rho }{\partial \textbf{x}}\frac{\partial \delta \textbf{x}}{%
\partial \textbf{x}} ,
\end{equation*}%
we obtain
\begin{equation*}
\delta \beta =2\,\left(\frac{\partial \delta \rho }{\partial \textbf{x}}-\frac{\partial \rho
}{\partial \textbf{x}}\frac{\partial \delta \textbf{x}}{\partial \textbf{x}}\right)\func{grad}\rho
\end{equation*}%
and%
\begin{equation*}
\delta \varepsilon =\varepsilon '_{\rho }\,\delta \rho +\varepsilon '
_{s}\,\delta s+\varepsilon '_{\beta } \,\delta \beta ,
\end{equation*}%
as well as eqs. (14.5) and (14.6) in \cite{Serrin}. For virtual
displacement  $\delta \textbf{x}$   null at the edge of $V$ and integration by
parts,  the integrals calculated on the edge of $V$ have a null
contribution; we obtain:
\begin{equation*}
\int_{V}(\rho\, \mathbf{a} -\func{div}\mathbf{\sigma} +\rho \func{grad}%
\Omega )\, \delta \textbf{x}\, dv=0,
\end{equation*}%
where $\mathbf{\sigma} $ is expressed by Eq. (\ref{3a}). Classical methods
of the calculus of variations   lead to Eq. (\ref{2}). It appears that a
single term $\lambda$ accounts for the effects of the second gradient deformation in the
equation of motion.  Scalar $\lambda$ depends on $\rho, s$ and $ \beta $ as $\varepsilon $ does. In a study of surface tension based on gas kinetic theory, Rocard
obtained the same expression as Eq. (\ref{3a}) for the stress tensor  but with $%
\lambda$ constant. If $\lambda$ is constant, specific internal energy $\varepsilon $ writes:%
\begin{equation*}
\varepsilon (\rho ,s,\beta )=\alpha (\rho ,s)+\frac{\lambda}{2\rho }\  \beta .
\end{equation*}%
The second gradient term $\displaystyle  {\lambda}\, \beta/(2\rho)$ is
simply added to the specific internal energy $\alpha (\rho ,s)$ of the classical compressible
fluid. The pressure of the compressible fluid is denoted $P \equiv \rho ^{2}\alpha _{\rho
}'  $  and the temperature is denoted $\Theta \equiv\alpha _{s}'  $.

Consequently,
\begin{equation*}
 p=P-\lambda\, \left(\frac{\beta
}{2}+\rho\, \Delta \rho\right).
\end{equation*}%
For the pressure $ P$, Rocard uses either the van der Waal  pressure
\begin{equation*}
P=\rho\, \frac{R\,\Theta }{1-b\rho }-a\rho ^{2}
\end{equation*}%
or other laws  of which he provides a comparison. It should be noted that
if $\lambda$ is constant, this implies that $\theta =\Theta $ and   there exits
a relationship, independent of the
second gradient terms, between $\theta ,\rho $ and $s$.

\subsection{Case of viscous fluids}

If the fluid is  endowed with viscosity, the equation of motion includes
not only the stress tensor $\mathbf{\mathbf{\sigma}} ,$ but also the
viscous stress tensor $\mathbf{\mathbf{\sigma}} _{v}$:
\begin{equation*}
\mathbf{\mathbf{\sigma}} _{v}=\nu\, \texttt{tr}\, \textbf{D} +2\,\mu\, \textbf{D},
\end{equation*}%
where $\textbf{D}$ is the deformation tensor,  symmetric gradient of the velocity
field \textbf{u}
\begin{equation*}
\textbf{D}=\frac{1}{2}\left\{\frac{\partial \textbf{u}}{\partial \textbf{x}}+\left(\frac{\partial \textbf{u}}{\partial \textbf{x}}
\right)^{T}\right\}.
\end{equation*}%
Of course  in second gradient theory, it would  be coherent   to add
 terms accounting for the influence of higher order derivatives of the
velocity field to the   viscous  stress tensor $\mathbf{\mathbf{\sigma}} _{v}$; this has not been done in the present case. Equation (2) is
modified by adding the virtual work of the forces of viscosity and we obtain:
\begin{equation*}
\rho\, \mathbf{a} =\func{div}(\mathbf{\mathbf{\sigma}} +\mathbf{\mathbf{%
\sigma}} _{v})-\rho \func{grad}\,\Omega .
\end{equation*}

\section{Equation of energy for   viscous fluids endowed with internal
capillarity}

Let $e$ be total volumic energy, $e\equiv\rho \displaystyle \left(\frac{1}{2}\, \textbf{u}^{2}+\varepsilon
+\Omega \right)$, $q$ the heat flux vector, $r$ the heat supply  and $h=\varepsilon +\displaystyle \frac{p}{\rho }$  the specific enthalpy.
We define the following quantities:

\begin{array}[t]{l}
\mathbf{M}=\rho\, \mathbf{\mathbf{a} }-\func{div}(\mathbf{\mathbf{\sigma}}
+\mathbf{\mathbf{\sigma}} _{v})+\rho\, \func{grad}\Omega, \\
B=\overset{\cdot }{\rho }+\rho \func{div}\mathbf{u}, \\
S=\rho\, \theta\, \overset{\cdot }{s}+\func{div}q-r-tr(\mathbf{\mathbf{\sigma}}
_{v}\textbf{D}), \\
E=\displaystyle\frac{\partial e}{\partial t}+\func{div}\left[(e-\mathbf{\mathbf{\sigma}} -%
\mathbf{\mathbf{\sigma}} _{v})\textbf{u}\right]-\func{div}(\lambda\,\overset{\cdot }{\rho }\func{%
grad}\rho )+\func{div}q-r-\rho \frac{\partial \Omega }{\partial t}\,.
\end{array}%

\emph{\textbf{Theorem}}:
For an internal energy written as in Eq. (1) and for any motion in the fluid, the
relation
\begin{equation}
E-\mathbf{M}\cdot \mathbf{u}-\left(\frac{1}{2}\,u^{2}+h+\Omega \right)B-S\equiv0 \label{energy}
\end{equation}%
is an identity.

\emph{Proof}: In the first member, the dissipative terms $q,r$ and $\mathbf{%
\mathbf{\sigma}} _{v}$ cancel out; this also is  the case for the extraneous force
potential and the inertia terms. After having replaced $\mathbf{\mathbf{%
\sigma}} ,p$ and $\theta $ by their respective values in Eqs (3 - 6), it remains to
  prove  that the terms from internal energy $\varepsilon $ also cancel
out. These terms include the following  expressions:

a) in $E:$ $\displaystyle \frac{\partial (\rho \varepsilon)}{\partial t}+\func{div}(\rho\,
\varepsilon\, \textbf{u})-\func{div}(\mathbf{\mathbf{\sigma}} \textbf{u})-\func{div}(\lambda\,\overset{%
\cdot }{\rho }\func{grad}\rho )$

$\qquad\qquad \equiv  \rho \overset{\cdot }{\varepsilon }+\varepsilon (\overset{\cdot }{\rho
}+\rho \func{div}\textbf{u})-\func{div}(\mathbf{\mathbf{\sigma}} )\textbf{u}-tr\left(\mathbf{%
\mathbf{\sigma}}\displaystyle \frac{\partial \textbf{u}}{\partial x}\right)-\func{div}(\lambda\,\overset{\cdot }{%
\rho }\func{grad}\rho )$,

or finally,

$\qquad\qquad \equiv\rho (\varepsilon _{s}^{\prime }\overset{\cdot }{s}+\varepsilon _{\rho
}^{\prime }\overset{\cdot }{\rho }+\varepsilon _{\beta }^{\prime }\overset{%
\cdot }{\beta })+\varepsilon (\overset{\cdot }{\rho }+\rho \func{div}\textbf{u})-%
\func{div}(\mathbf{\mathbf{\sigma}} )\textbf{u}+p\func{div}\textbf{u}$

$\qquad\qquad + \lambda\,(\func{grad}\rho )^{T}\displaystyle
\frac{\partial \textbf{u}}{\partial x}\func{grad}\rho -\func{div}(\lambda\,\overset{\cdot }{%
\rho }\func{grad}\rho )$;

b) in $\mathbf{M\cdot \textbf{u}}$ : $-\func{div}(\mathbf{\mathbf{\sigma}} )\mathbf{u}
$;

c) in $\displaystyle\left(\frac{1}{2}\,\textbf{u}^{2}+h+\Omega \right)B$ : $\displaystyle\left(\varepsilon +\frac{p}{\rho }\right)(%
\overset{\cdot }{\rho }+\rho \func{div}\textbf{u})$;

d) in $S$ : $\rho\, \varepsilon _{s}^{\prime }\,\overset{\cdot }{s}$.

So, the first member of Eq. (\ref{energy}) writes:
\begin{equation*}
\rho\, \varepsilon _{\rho }^{\prime }\,\overset{\cdot }{\rho }+\frac{\lambda}{2}\,
\overset{\cdot }{\beta }-\frac{p}{\rho }\,\overset{\cdot }{\rho }+\lambda\,(\func{grad}%
\rho )^{T}\frac{\partial V}{\partial x}\func{grad}\rho -\func{div}(\lambda\,\overset{%
\cdot }{\rho }\func{grad}\rho) .
\end{equation*}%
Finally, by taking Eq. (5) into account, we obtain that
\begin{equation*}
\lambda\,(\func{grad}\rho )^{T}\frac{d(\func{grad}\rho)}{dt} +\overset{\cdot }{\rho }\,
\func{div} \lambda\,(\func{grad}\rho )+\lambda\,(\func{grad}\rho )^{T}\frac{\partial \textbf{u}}{%
\partial \textbf{x}}\func{grad}\rho -\func{div}(\lambda\,\overset{\cdot }{\rho }\func{grad}%
\rho )
\end{equation*}%
is identically null.

\emph{\textbf{Corollary 1}}:
For a conservative motion of a fluid endowed with internal capillarity, the
conservation of entropy along the trajectories is equivalent to
\begin{equation}
\frac{\partial e}{\partial t}+\func{div}[(e-\mathbf{\mathbf{\sigma}} )\textbf{u}]-%
\func{div}(\lambda\,\overset{\cdot }{\rho }\func{grad}\rho )-\rho\; \frac{\partial
\Omega }{\partial t}=0 , \label{corolaire1}
\end{equation}%
which is  the equation of conservation of energy. This is
derived from Eq. (\ref{energy}) when $\mathbf{\mathbf{\sigma}} _{v}=0,\func{div}q=r=0$ and $
 M=0, B=0, S=0$. It leads us to add the additional term $\func{div}(\lambda\,
\overset{\cdot }{\rho }\func{grad}\rho )$ to the equation of energy. The
vector $ \lambda\,\overset{\cdot }{\rho }\func{grad}\rho $ has the dimension of a
heat flux vector and even occurs in the conservative case.

\emph{\textbf{Corollary 2}}:
For a motion of a viscous fluid endowed with internal capillarity, the energy
equation
\begin{equation*}
\frac{\partial e}{\partial t}+\func{div}[(e-\mathbf{\mathbf{\sigma}} -%
\mathbf{\mathbf{\sigma}} _{v})\textbf{u}]-\func{div}(\lambda\,\overset{\cdot }{\rho }\func{%
grad}\rho )+\func{div}q-r-\rho \frac{\partial \Omega }{\partial t}=0
\end{equation*}%
is equivalent to the entropy equation
\begin{equation}
\rho \theta \overset{\cdot }{s}+\func{div}q-r-tr(\mathbf{\mathbf{\sigma}}
_{v}\textbf{D})=0 .\label{planck0}
\end{equation}%
Equation (\ref{planck0}) corresponds to the classical version of the entropy variation
expressed by the function of dissipation of viscous stresses
$
\psi =tr(\mathbf{\mathbf{\sigma}} _{v}\textbf{D}).
$

\section{The Planck inequality and the Clausius - Duhem inequality}

For any motion in a viscous fluid endowed with capillarity, it has been
assumed that
\begin{equation}
\rho\, \theta \,\overset{\cdot }{s}+\func{div}q-r\geqslant 0\, . \label{planck}
\end{equation}%
Inequality (\ref{planck}) represents the Planck inequality \cite{Truesdell}. Let generally write the Fourier
principle as
\begin{equation*}
q\cdot \func{grad}\theta \leqslant 0 \,.
\end{equation*}
Then, the Clausius-Duhem inequality can be directly deduced as
\begin{equation*}
\rho\, \overset{\cdot }{s}+\func{div}\frac{q}{\theta }-\frac{r}{\theta }%
\geqslant 0 \,.
\end{equation*}%
When in motion, the fluid is endowed with internal capillarity, the law of thermodynamics leads to the existence of a heat flux
vector.
This heat flux vector introduces an additional term into the classical
equation of energy, even if the fluid is non-viscous. It is possible to extend these results   to
continuous media with an internal energy that contains
gradient terms of deformation.

\section{Transformation of the motion equations for   fluids endowed with
internal capillarity}

In this section, we rewrite Eq. (2) in other forms.

\subsection{Case $\lambda$ constant}

Let us note
$
\omega =\Omega -\lambda\,\Delta \rho .
$
Then, Eq. (2) can be written
\begin{equation}
\rho\, \mathbf{a} +\func{grad}\emph{P}+\rho \func{grad}\, \omega =0\,.\label{motion2}
\end{equation}
This is a  representation of a perfect fluid where $\emph{P}$\, is the
van der Waals pressure;  term $\omega $ contains all   capillarity
terms. \newline
Equations (3) and (5) yield
\begin{equation*}
\mathbf{\mathbf{\sigma}} _{ij}=-\emph{P}\,\delta _{ij}+\lambda\,\left\{\left(\frac{1}{2}\ \rho
_{,k} \rho _{,k}+\rho \rho ,_{kk}\right)\delta _{ij}-\lambda\,\rho _{,i} \rho _{,j}\right\}.
\end{equation*}%
Then,%
\begin{equation*}
\mathbf{\mathbf{\sigma}} _{ij,j}=-\emph{P}_{,i}+\lambda\,\rho \rho _{,ijj}
\end{equation*}%
i.e.
\begin{equation*}
\func{div}\mathbf{\mathbf{\sigma}} =-\func{grad}\emph{P}+\lambda\,\rho \func{grad}%
(\Delta \rho )\,.
\end{equation*}%
We can note that for viscous fluids, Eq. (\ref{motion2}) writes
\begin{equation}
\rho\, \mathbf{a} +\func{grad}\emph{P}+\rho \func{grad} (\Omega -\lambda\,\Delta
\rho )-\func{div}\mathbf{\mathbf{\sigma}} _{v}=0\,. \label{vicous motions}
\end{equation}

\subsection{Thermodynamic form of the equation of motion}

Commonly - and not only when $\lambda$ is constant - the equation of motion (\ref{2}) can
be written in the following form
\begin{equation}
\mathbf{a} =\theta \func{grad}s-\func{grad}(h+\Omega ),\label{thermotion}
\end{equation}
which is the \emph{thermodynamic form} of the equation of motion. In the case
without capillarity ($\varepsilon =\alpha (\rho ,s)$, $\varepsilon ' _{\beta
} =0)$, Eq. (\ref{thermotion}) is well-known (\cite{Serrin}, p. 171).
The equation remains valid for fluids endowed with internal capillarity.

The verification simply consists on writing out the second member of
Eq. (\ref{thermotion}). Let us denote%
\begin{equation*}
\begin{array}{ccc}
A=\theta \func{grad}s-\func{grad}h &\quad \text{i.e.}\quad & A_{i}=\theta s_{,i}-h_{,i}\,.
\end{array}%
\end{equation*}%
According to Eq. (\ref{temperature}), we get:
$\displaystyle
A_{i}=\varepsilon _{s}^{\prime }s_{,i}- \left(\varepsilon +\frac{p}{%
\rho }\right)_{,i} .
$
Then,
\begin{equation}
A_{i}=\varepsilon _{s}^{\prime }s_{,i}-\left\{\varepsilon _{s}^{\prime
}s_{,i}+\varepsilon _{\rho }^{\prime }\,\rho _{,i}+\varepsilon _{\beta
}^{\prime }\left (\rho _{,k}\rho _{,k}\right)_{,i}-\frac{p}{\rho
{{}^2}%
}\,\rho _{,i}-\frac{1}{\rho }\,p_{,i}\right\}.\label{intermediate}
\end{equation}%
Using Eq. (\ref{sigma}), Eq. (\ref{intermediate})  writes $\displaystyle  \mathbf{\mathbf{\sigma}}
_{ij,j}/ \rho$ and consequently
$
A=\func{div}\mathbf{\mathbf{\sigma}} /\rho,
$
which proves Eq. ({\ref{thermotion}).

\section{Generalized Kelvin theorems}

Let $J$ be the circulation of the velocity vector along a closed fluid curve
$C$ convected by the fluid flow
\begin{equation*}
J=\oint_{C}\textbf{u}^{T} d\textbf{x},
\end{equation*}%
and \cite{Serrin},
\begin{equation*}
\frac{dJ}{dt}=\oint_{C}\mathbf{a} ^{T} d\textbf{x} .
\end{equation*}%
Equation (\ref{thermotion}) implies $\displaystyle  \oint_{C}\mathbf{a} ^{T} d\textbf{x}=\oint_{C}\theta\, ds$
and consequently yields the following theorems which are valid for fluids
endowed with internal capillarity:

\emph{\textbf{Theorem 1}}: The velocity circulation along a closed  isentropic fluid
curve is constant.

\emph{\textbf{Corollary 1}} : For a homentropic flow [27], the velocity circulation along a
closed fluid curve is constant.

\emph{\textbf{Theorem 2}}: The velocity circulation along a closed isothermal fluid
curve is constant.

\emph{\textbf{Corollary 2}} : For an isothermal flow, the velocity circulation along a
closed fluid curve is constant.

\section{Potential equations for conservative fluids and classification of   motions}

The results of \cite{Casal 1,Casal 5,Seliger,Gouin 2} can be applied to conservative motions of fluids
endowed with internal capillarity. To any motion of  fluids endowed with internal capillarity, there
correspond  scalar potentials $\varphi ,\psi ,\tau $ and $\chi $  verifying
\begin{equation}
\left\{
\begin{array}{l}
\quad\overset{\cdot }{\varphi }=\displaystyle\frac{1}{2}\ \textbf{u}^{2}-h-\Omega ,   \\
\quad\overset{\cdot }{%
\tau }=0,\\
\quad\overset{\cdot }{\psi }=\theta ,\\ \quad   {\overset{\cdot }{\chi }=0},
\\
\quad\overset{\cdot }{s}=0 .
\end{array}
\right.
\label{tableau}
\end{equation}
The fluid velocity is given by
\begin{equation}
\textbf{u}=\func{grad}\varphi +\psi \func{grad}s+\tau \func{grad}\chi ,\label{potentialequ.}
\end{equation}%
together with
\begin{equation*}
\frac{\partial \rho }{\partial t}+\func{div}(\rho\, \textbf{u})=0.
\end{equation*}

The   equations can be used to classify motions in the same way as in the
case of non-capillary    perfect fluids [27,30].

\subsection{Homentropic motions}

This case corresponds  to  $s$   constant in all the fluid. Eq. (\ref{potentialequ.}) writes
\begin{equation*}
\textbf{u}=\func{grad}\varphi+\tau \func{grad}\chi ,
\end{equation*}
which leads to the Cauchy  theorem \cite{Serrin}
\begin{equation*}
\frac{d}{dt}\left(\frac{\func{rot} \textbf{u}}{\rho }\right)=\frac{\partial \textbf{u}}{\partial \textbf{x}}\frac{\func{rot} \textbf{u}}{%
\rho } .
\end{equation*}

\subsection{Oligotropic motions}

Equation (\ref{potentialequ.})  writes
\begin{equation*}
\textbf{u}=\func{grad}\varphi +\psi \func{grad}s .
\end{equation*}
The flows verify
the relation $\func{rot} \textbf{u}\,_\cdot \func{grad}s=0$. Surfaces with equal entropy are
eddy surfaces. The velocity circulation along a closed  isentropic fluid curve
is null.

\emph{Remark:} the form of Eqs. (\ref{motion2}) and (\ref{thermotion}) allows to generalize the results
of \cite{Gouin 3}.

\section{Generalized Crocco -Vazsonyi equation}

The equation of energy (\ref{corolaire1}) can be also written
\begin{equation}
\frac{\partial e}{\partial t}+\func{div}[\,(e+p)\,\textbf{u}\,]-\func{div}\left(\lambda\,\frac{\partial
\rho }{\partial t}\func{grad}\rho \right)-\rho \frac{\partial \Omega }{\partial t}=0 ,\label{energie1}
\end{equation}
with $\displaystyle e+p=\rho \left(\frac{1}{2}\,\textbf{u}^2
+h+\Omega \right)$ and denoting $\displaystyle H=\frac{1}{2}\,\textbf{u}^2
+h+\Omega $ , Eq. (\ref{energie1}) writes
\begin{equation}
\frac{\partial \rho H}{\partial t}+\func{div}\rho H \textbf{u}=\rho \frac{\partial
\Omega }{\partial t}+\frac{\partial p}{\partial t}+\func{div}\left(\lambda\,\frac{%
\partial \rho }{\partial t}\func{grad}\rho \right).\label{energy3}
\end{equation}%
The first member is equal to $\rho \overset{\cdot }{H}+H(\overset{\cdot }{%
\rho }+\rho \func{div}\textbf{u})$. Because of mass conservation, the
definitive form of the equation of energy writes
\begin{equation*}
\rho \overset{\cdot }{H}=\rho \frac{\partial \Omega }{\partial t}+\frac{%
\partial p}{\partial t}+\func{div}\left(\lambda\,\frac{\partial \rho }{\partial t}\func{%
grad}\rho \right).
\end{equation*}%
By taking account of the identity
\begin{equation*}
\mathbf{a} =\frac{\partial \textbf{u}}{\partial t}+\func{rot} \wedge\, \textbf{u}+\func{grad}\left(%
\frac{1}{2}\,\textbf{u}^2\right),
\end{equation*}
Equation (\ref{thermotion}) can be written
\begin{equation}
\frac{\partial \textbf{u}}{\partial t}+\func{rot} \textbf{u}\wedge\, \textbf{u}=\theta \func{grad}s-\func{grad}H .\label{motion3}
\end{equation}
If the motion is a steady flow,   Eq. (\ref{motion3}) makes it possible to conclude that $
H $ is constant along the stream lines $(\overset{\cdot }{H}=0)$   and Eq.
(\ref{energy3}) shows that the divergence of the partial energy flux $\rho H \textbf{u}$ is null. Equation
(\ref{motion3})   leads to the Crocco-Vazsonyi equation generalized to fluids
endowed with internal capillarity; in stationary motion,
\begin{equation*}
\func{rot} \textbf{u}\wedge \textbf{u}=\theta \func{grad}s-\func{grad}H,
\end{equation*}%
where  $s$ and $H$ are constant along every stream
line.

\emph{Remark}: Equations (\ref{tableau}) show that along each trajectory, and consequently across
interfaces, $\tau $, $\chi$ and the entropy $s$  are constant scalars. They represent first integrals of the motion. Other integrals
can obviously be found as the Kelvin integrals
\begin{equation*}
J=\int_{\Gamma} \tau\, (\func{grad}\chi )^{T} d\textbf{x}\equiv\int_{\Gamma}\tau \,d\chi
\end{equation*}%
are constant along any   fluid curve  $\Gamma$. Thanks to the
Noether theorem, any law of conservation can be represented with an
invariance group \cite{Gouin 5}. The law of conservation expressed with the Kelvin theorems
corresponding to isentropic fluid curves is related to the group of
permutations associated with   particles of equal entropy; the group
keeps the equations of motion invariant for both classical perfect fluid and
fluid endowed with internal capillarity, but also for any perfect fluid
endowed with an internal energy depending on $\rho , s$  and their
gradients of any order. It is even tempting to define a general perfect
fluid by identifying it with an invariant group or with a
continuous medium whose motions verify the Kelvin theorems \cite{Gouin 4}.

\section{Dynamic surface tension of liquid-vapor interfaces}

Far from  the critical conditions, the thickness of a
liquid-vapor interface is very small \cite{Rocard,Rowlinson}. Outside the capillary layer,
density and its spatial derivatives have smooth variations. The density in
each phase is reached at points located within the immediate vicinity of
the layer. In our study, bubbles and drops of  size of a few
molecular beams have not been considered. Surfaces of equal density
materializing the interface are  parallel surfaces  and are
used to define a system of orthogonal coordinates \cite{Valiron}. Notations are those
used in \cite{Germain 2}. The subscript 3 refers to the direction normal to surfaces
of equal density. Let us denote by $\textbf{e}_{3}$
the unit vector of this index oriented in the increasing density direction. The mass flow through the interface is
assumed to be null such that surfaces of equal density
are material surfaces. In the capillary layer,
\begin{equation}
\func{div}\textbf{u}=0. \label{incompress}
\end{equation}
Extraneous forces being neglected,   Eq. (\ref{incompress})
represents velocities compatible with the interface. Equation (\ref{vicous motions}) yields:
\begin{equation}
\rho\, {\mathbf{a}} _3+\frac{1}{h_{3}}\frac{\partial {P}}{\partial x_{3}}%
=\lambda\,\rho \frac{1}{h_{3}}\frac{\partial {  \Delta \rho }}{\partial x_{3}}+2\mu\, (%
\func{div}\textbf{D})_{3},\label{tangential}
\end{equation}%
\begin{equation*}
\rho\, \mathbf{a} _{tg}+\text{grad}_{tg} {P}=\lambda\,\rho\, {\text{grad}_{tg}} {%
\Delta \rho }+2\mu\, (\func{div}\textbf{D})_{tg},
\end{equation*}%
where the subscript $tg$ denotes the tangent component to the surfaces of
equal density.  The different
linear sizes of the interface should be taken in consideration. The
capillary layer is measured in Angstr\" om  and the surface curvature radii
are of non-molecular dimensions. The deduced relations result from a limit
analysis where the parameter related to the thickness of the capillary
layer tends to zero.

The subscripts $v$ and $l$ designating the vapor and the liquid bulks,
respectively,  integration of
Eq. (\ref{tangential}) along the third coordinate line  yields, when we assume that $\mathbf{a} _{3}$ is bounded through the interface such that its integral is negligible,
\begin{equation}
\emph{P - P}_{v}=\lambda\,\rho\, \Delta \rho -\lambda\,\int_{x_{3}^{v}}^{x_{3}}\Delta \rho\, \frac{%
\partial \rho }{\partial x_{3}}\, dx_{3}+2 \int_{x_{3}^{v}}^{x_{3}}\mu \,(\func{%
div}D)_{3} \, h_{3}\, dx_{3}\, .\label{tangint}
\end{equation}%
The partial derivatives of velocity with respect to coordinates $x_{1}$ and $%
x_{2}$ are assumed to be bounded. Taking account of Eq. (\ref{incompress}), the last term in
Eq. (\ref{tangint}) is negligible. Furthemore, we have the relation:
\begin{equation}
\Delta \rho =-\frac{2}{R_{m}}\frac{1}{h_{3}}\frac{\partial \rho }{\partial
x_{3}}+\frac{1}{h_{3}}\left(\frac{1}{h_{3}}\frac{\partial \rho }{\partial x_{3}}
\right)_{,3} ,\label{laplacien}
\end{equation}%
where $R_{m}$ denotes the mean curvature of the surfaces of equal density in
the capillary layer, oriented by $\textbf{e}_{3}$ \cite{Germain 1}. It follows:
\begin{equation}
\emph{P - P}_{v}=\lambda\,\left\{\rho \Delta \rho -\frac{1}{2}\,(\func{grad}\rho )%
^2\right\}+2\frac{\lambda\,}{R_{m}}\int_{x_{3}^{v}}^{x_{3}}(\func{grad}\rho )^2\,
h_{3}\, dx_{3}.\label{pressuredev}
\end{equation}%
Denoting $dn\equiv h_{3}\, dx_{3},$ we finally obtain:
\begin{equation*}
\begin{array}{lllll}
\displaystyle \emph{P - P}_{v}=2\,\frac{H}{R_{m}} &  & \text{where} &  & \displaystyle H=\lambda\,
\int_{n_{v}}^{n_{l}}(\func{grad}\rho )^2
dn,
\end{array}%
\end{equation*}%
which expresses the Laplace  equation  for motions.  We interpret $H$ as the dynamical  surface tension.

\emph{Remark: Interpretation of the surface tension at equilibrium}.\newline  The eigenvalues of the internal capillary stress tensor are deduced from Eq. (\ref{3a}):
\\
$\lambda _{1}=-p\, +\, \lambda\,(\func{grad}\rho )^2$ is the eigenvalue associated with the plane perpendicular to $\func{grad}
\rho $.
\\
$\lambda _{2\text{ }}=-p$ is the eigenvalue associated with the direction $%
\func{grad}\rho .$

In the system of curvilinear coordinates related to the interface, the
stress tensor writes%
\begin{equation*}
\mathbf{\mathbf{\sigma}} =%
\begin{bmatrix}
\lambda _{1} & 0 & 0 \\
0 & \lambda _{1} & 0 \\
0 & 0 & \lambda _{2}%
\end{bmatrix}%
.
\end{equation*}%
The equilibrium equation for the plane interface is drawn from Eq. (\ref{2});
neglecting extraneous forces we get:
\begin{equation*}
\lambda _{2}=-\emph{P}_{_0},
\end{equation*}%
where $\emph{P}_{_0}$ denotes the pressure in the liquid and vapor
bulks. The force per unit length along the edge of the interface (see Fig. 1)  is:
\begin{equation*}
F=\int_{x_{3}^{v}}^{x_{3}^{l}}\lambda _{1} h_{3}\, dx_{3}=-\emph{P}_{_0}\,
L+\int_{x_{3}^{v}}^{x_{3}^{l}}\lambda\,(\func{grad}\rho )^2\,
h_{3}\, dx_{3} ,
\end{equation*}%
where $L$ denotes the capillary layer thickness.
\begin{figure}[h]
\begin{center}
\includegraphics[width=9cm]{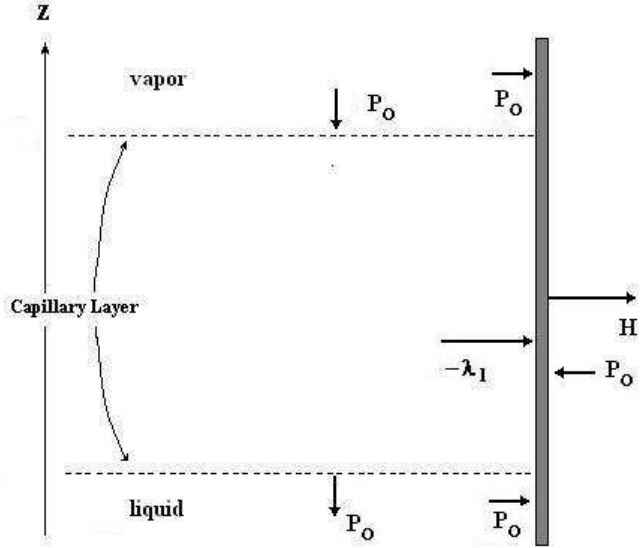}
\end{center}
\caption{\emph{Interpretation of the surface tension in liquid-vapor interfaces as a layer of fluid endowed with internal capillarity.}}\label{fig1}
\end{figure}
 Obviously, $\emph{P}_{_0}\, L$
is negligible.  Let us denote
\begin{equation*}
H=\int_{x_{3}^{v}}^{x_{3}^{l}}\lambda\,(\func{grad}\rho )^2\,
h_{3}\, dx_{3},
\end{equation*}%
the force per unit length; then $H$ represents the surface tension of the
plane interface at equilibrium.

\section{Practical calculation of the surface tension}

By integrating Eq. (\ref{tangential}) on the third coordinate line, the viscosity term
being assumed negligible, we obtain
\begin{equation}
\lambda\,\Delta \rho =\frac{\partial }{\partial \rho }\left(\rho \int_{\rho _{v}}^{\rho }%
\frac{\emph{P - P}_{v}}{\rho^2}\, d\rho \right).\label{laplacien2}
\end{equation}%
Let us notice that $\displaystyle\frac{\partial \emph{P}}{\partial \theta }\frac{\partial
\theta }{\partial x_{3}}$ is negligible with respect to $\displaystyle\frac{\partial
\emph{P}}{\partial \rho }\frac{\partial \rho }{\partial x_{3}}.$

Taking Eq. (\ref{laplacien}) into account, a further integration on the third coordinate
leads to
\begin{equation}
\frac{\lambda}{2}(\func{grad}\rho )
^2
=2\frac{\lambda}{R_{m}}\int_{x_{3}^{v}}^{x_{3}^{{}}}(\func{grad}\rho )^2
h_{3}\, dx_{3}+\rho \int_{\rho _{v}}^{\rho }\frac{\emph{P - P}_{v}}{\rho
^2}\, d\rho\, .\label{densities}
\end{equation}%
By denoting $x_{3}^{i}$ the third coordinate of the surface of equal density   $%
\rho =\rho _{i}$, then  for $x_{3}\in \lbrack x_{3}^{v},x_{3}^{i}]$, the term $\ \displaystyle 2%
\frac{\lambda}{R_{m}}\int_{x_{3}^{v}}^{x_{3}^{{}}}(\func{grad}\rho )^2
h_{3}\, dx_{3}\ $ is negligible with respect to $\ \displaystyle \frac{\lambda}{2}\,(\func{grad}\rho )^2
$; that is the case  for $\displaystyle \rho _{i}=\frac{1}{2}\,(\rho _{v}+\rho _{l})$. For $\rho \in \lbrack \rho _{v},\rho _{i}]$,\ \ Eq. (\ref{densities}) writes
\begin{equation*}
\frac{\lambda}{2}(\func{grad}\rho )^2
=\rho \int_{\rho _{v}}^{\rho }\frac{\emph{P - P}_{v}}{\rho
^2
} d\rho.
\end{equation*}
We obtain the same result for $\rho \in \lbrack \rho _{i},\rho _{l}]$ and
finally,
\begin{equation}
H=\sqrt{2\lambda\,}\left(\int_{\rho _{v}}^{\rho _{i}}\sqrt{u\int_{\rho _{v}}^{u}\frac{%
\emph{P - P}_{v}}{\rho
^2
} d\rho }\  du+\int_{\rho _{i}}^{\rho _{l}}\sqrt{u\int_{\rho _{i}}^{u}\frac{%
\emph{P - P}_{l}}{\rho
^2
} d\rho\ }du \right) .\label{surtension}
\end{equation}%
Relation (\ref{surtension}) expresses the surface tension of a liquid-vapor interface for
motions compatible with the interface. Here $\emph{P}$ is not only a function
of density but also temperature and varies according to the location of the
point within the interface. The viscosity has explicitly disappeared in the
expression of $H$ (the expression of $H$ does not explicitly   take the viscosity  into
account). In the case of plane interface in isothermal
equilibrium, the expression given by Rocard \cite{Rocard} is found again. The value of
the internal capillarity constant can   be numerically calculated on
the basis of experimental values of $H$ and  expressions of $P$.
By injecting the value of $\lambda$ in Eq. (\ref{surtension}), we can calculate surface tension
for any dynamic temperature distribution.

\emph{Remark:} Eq. (\ref{laplacien2}) yields
\begin{equation*}
\lambda\,\Delta \rho =\frac{\emph{P}}{\rho
}-\frac{\emph{P}_{v}}{\rho _{v}}+\int_{\rho _{v}}^{\rho }\frac{\emph{P}}{%
\rho
^2
} d\rho
\end{equation*}%
and we deduce
\begin{equation*}
\int_{\rho _{v}}^{\rho _{l}}\frac{\emph{P}} {\rho
^2
}\, d\rho =\frac{\emph{P}_{v}}{\rho _{v}}-\frac{\emph{P}_{l}}{\rho _{l}}
\end{equation*}
which represents an invariant  corresponding to motions compatible
with the interface. This invariant leads to the equations (4-11) of \cite{Aifantis} demonstrated
in the specific case of an interface in isothermal equilibrium. The case of
plane interface    leads to the Maxwell equal area rule \cite{Rocard,Aifantis}.

\section{Marangoni effect in liquid-vapor interfaces}

Let us denote
\begin{equation*}
H_{j}\left(x_{3}\right)=\int_{x_{3}^{j}}^{x_{3}^{{}}}(\func{grad}\rho ) ^2
h_{3}\,dx_{3},\qquad \rm{where}\quad j\in \{\emph{v},\emph{l}\}.
\end{equation*}%
By using calculations as in Section 10, for example for $\ x_{3}\in \lbrack
x_{3}^{v},x_{3}^{l}]$, we obtain
\begin{equation*}
H_{v}(x_{3})=\sqrt{2\lambda\,}\left(\int_{\rho _{v}}^{\rho _{{}}}\sqrt{u\int_{\rho
_{v}}^{u}\frac{\emph{P - P}_{v}}{\rho
^2
}d\rho }\ du\right) .
\end{equation*}%
 By injecting  Eq. (\ref{pressuredev})   into Eq. (\ref{laplacien2}), we obtain:
\begin{equation}
\rho\, \mathbf{a }_{tg}=\frac{\lambda}{2}\,\text{grad}_{tg}(\func{grad}\rho )%
^2
-\text{grad}_{tg}\left(\frac{2}{R_{m}}H_{j}(x_{3})\right)-\text{grad}_{tg}\emph{P}%
_{j}+2\mu (\func{div}D)_{tg}   .\label{tangential2}
\end{equation}%
Equation (\ref{incompress}) and conditions on partial derivatives of velocity imply:
\begin{equation*}
\int_{x_{3}^{v}}^{x_{3}^{l}}(\func{div}%
D)_{tg}h_{3}\,dx_{3}=[De_{3}]_{x_{3}^{v}}^{x_{3}^{l}}\,.
\end{equation*}%
Integration of Eq. (\ref{tangential2}) yields:
\begin{equation*}
\frac{\lambda}{2}\int_{x_{3}^{v}}^{x_{3}^{l}} \text{grad}_{tg}(\func{grad}\rho )^2
h_{3}\,dx_{3}+2[\mu De_{3}]_{x_{3}^{v}}^{x_{3}^{l}}=0\,.
\end{equation*}%
Taking account of
\begin{equation*}
\frac{\lambda}{2}\int_{x_{3}^{v}}^{x_{3}^{l}} \text{grad}_{tg}(\func{grad}\rho ) ^2
h_{3}\,dx_{3}=\text{grad}_{tg}\left\{\lambda \int_{x_{3}^{v}}^{x_{3}^{l}}(\func{grad}\rho ) ^2
h_{3}\,dx_{3}\right\},
\end{equation*}%
we can conclude%
\begin{equation*}
\text{grad}_{tg}H+2[\mu De_{3}]_{v}^{l}=0\,.
\end{equation*}%
If the viscosity stress tensor of the vapor bulk is supposed to be negligible, we get:
\begin{equation*}
\text{grad}_{tg}H+2[\mu D^{l}e_{3}]=0\,.
\end{equation*}%
This is the usual Marangoni condition for free boundary
problems \cite{Sen}. In the limit case when the viscosity coefficient is null (as for
superfluid helium), the problem must be proposed in another way: the
momentum associated with the interface can no longer be neglected and other
physical effects must be taken into account \cite{Landau}. We must note there exist
other phenomenal presentations of the Marangoni effect (for example \cite{Scriven}),
but to our knowledge, they all consider the interface as a discontinuous
surface of the fluid medium. The calculations performed in the present text
do not call upon the use of any linear approximations, and only take account
of the various physical quantities describing liquid-vapor interfaces while
working far from the critical temperature. The use of second gradient theory
for representing interfaces has, of course, been considered by several
authors. In \cite{Dunn,Slemrod,Slemrod 2,Dunn 2},  their source is found to be in free energy which is
more directly useable in case of isothermal flows.

\section{Conclusion and further developments}

As we have seen in this revisited version of \cite{Gouin 0}, the second gradient theory, conceptually more straightforward than the Laplace theory can be used to build a theory of capillarity. Such a theory is able to take  account of systems in which fluid interfaces are present. The \emph{internal capillarity} associated with  fluids  is the simplest case.
A mathematical limit analysis associated with the thickness of the interface when the size of the layer goes toward zero and the behavior of the layer between fluid phases yield the model of material surfaces.
Such a theory is able to calculate the superficial tension as well in the case of thin interfaces as thick interfaces.
The static model in continuum mechanics of second gradient theory is extended to dynamics. The equation of motion is able to induce a stress tensor. In the fluid case, the theory does not lead to an isotropic stress tensor.

\emph{\textbf{\footnotesize Many developments have been done since paper \cite{Gouin 0} by the author:}}

It is possible to obtain the radius of nucleation of microscopic drops and bubbles and to develop a macroscopic theory as the Laplace theory of capillarity for curved interfaces \cite{Isola}. The stability of interfaces is investigated with differential or partial derivative equations \cite{Gouin Slemrod}. Classification of fluids endowed with internal capillarity is the extension of the classical fluid one's \cite{Casal 7}.\newline
In fact, contact forces are  of a different nature than the ones associated with the Cauchy stress tensor. Classical conditions with the tetrahedron construction due to Cauchy are not efficient to study the non-linear behavior of fluids endowed with internal capillarity. We deduced contact forces concentrated on edges representing the boundaries of surfaces of separation. For example, such conditions are necessary to study the stability of thin films in contact with solid walls and the connection with the mean field molecular theories \cite{Gouin 8,Gouin 9,Gouin 16,Gouin 14,Gouin 13}. \newline
The model  dynamically interprets phenomena in the vicinity of the critical point and authorizes an investigation - at least qualitative - of the dynamic change of phases between bulks in fluids or fluids mixtures \cite{Gouin 7,Gouin 11}. New mathematical equations associated with the hyperbolicity in continuous media can be considered in second gradient theory and consequently in internal capillarity \cite{Gouin 10,Gouin 17}.\newline
More recently, we must notice that fluids endowed with internal capillarity are able to model  fluid layers at a scale of some nanometers and to recognize the disjoining pressure concept for very thin liquid films \cite{Gouin 13}. The model can be applied in vegetal biology to interpret the ascent of the crude sap in very high trees as sequoia and giant eucalyptus \cite{Gouin 12,Gouin 15}

\end{document}